\documentclass[aps,twocolumn,showpacs]{revtex4}

\usepackage{graphicx}

\begin{document}
\title{
The simplest quantum model supporting the Kibble-Zurek mechanism 
of topological defect production: Landau-Zener transitions 
from a new perspective
}
\author{Bogdan Damski}
\affiliation{
 Institut f\"ur Theoretische Physik, Universit\"at Hannover, Appelstr. 2,
 D-30167
 Hannover, Germany\\
Instytut Fizyki, Uniwersytet
Jagiello\'nski, Reymonta 4, PL-30-059 Krak\'ow, Poland}
\begin{abstract}
It is shown that dynamics of the Landau-Zener model can be 
accurately described in terms of the Kibble-Zurek theory 
of the topological defect production in nonequilibrium 
phase transitions. The simplest 
quantum model exhibiting  the Kibble-Zurek mechanism 
is presented. A new intuitive description of Landau-Zener 
dynamics is found.
\end{abstract}
\pacs{03.65.-w,03.75.Lm,32.80.Bx,05.70.Fh}
\maketitle

In this Letter we present a  successful
combination of the Kibble-Zurek (KZ) \cite{kibble,zurek} theory of 
topological defect production 
and  quantum theory of the Landau-Zener (LZ) model \cite{LZ}. 
Both theories play a prominent role in contemporary 
physics. The KZ theory  predicts production of topological defects 
(vortices, strings) 
in the course of nonequilibrium phase transitions.
This prediction applies to phase transitions in 
liquid $^4$He and $^3$He, liquid crystals,
superconductors, ultracold atoms in optical lattices \cite{kz,siec}, 
and even to cosmological phase transitions in 
the early Universe \cite{kibble,zurek}.
The Landau-Zener theory
has even broader
applications. It has already  become a standard tool
in quantum optics, atomic and 
molecular physics, and
solid state physics. 
The list of important physical 
systems governed by the LZ model grows.
For instance, recent investigations point out that 
the smallest quantum magnets, Fe$_8$ clusters cooled 
below $0.36$K, are successfully described by the LZ model \cite{magnets}.

This Letter constructs the simplest quantum model
whose  dynamics remarkably resembles  dynamics of 
topological defect production in nonequilibrium 
second order phase transitions. The model is built on the basis
of LZ theory and allows us to study the KZ mechanism of topological
defect production in a truly  quantum case. 
Such a quantum insight into the KZ theory 
was up to now inaccessible except for the recent study of
KZ theory in optical lattices filled with ultracold atoms \cite{siec}.
In addition, we present  a 
simple, intuitive, and accurate 
description of   LZ model dynamics.

\begin{figure}[h]
\includegraphics[angle=-0,scale=0.25, clip=true]{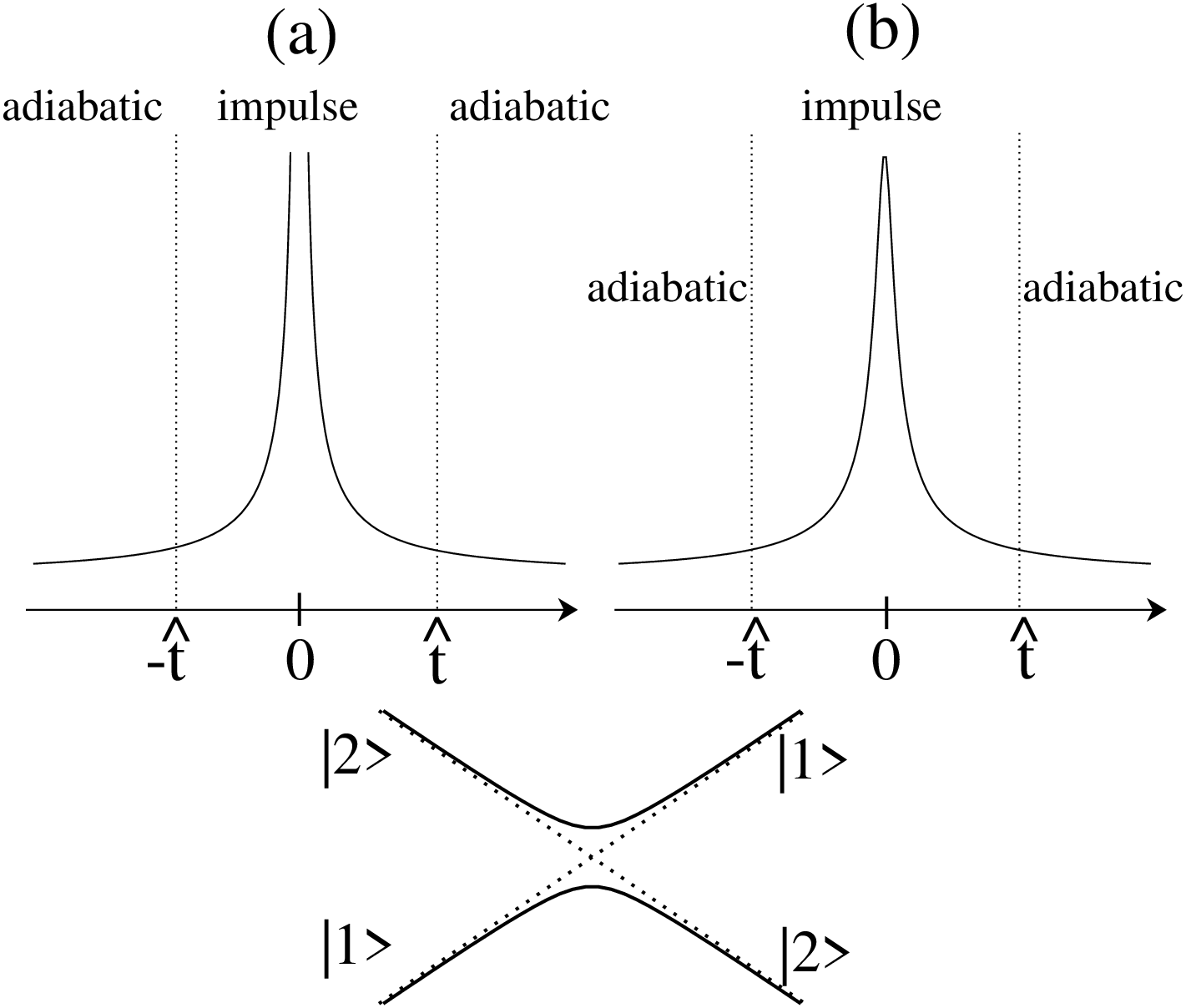}
\caption{Plot (a):  relaxation time scale $\tau$ from the KZ theory.
Plot (b): inverse of the gap in the LZ model (\ref{eq1}). 
Lower plot: energy levels of the Hamiltonian (\ref{eq1}); 
dotted line: $\omega_0=0$ case.}
\label{fig1}
\end{figure}

For the rest of 
the  Letter it is essential to introduce briefly the KZ theory.
Consider a pressure quench that drives liquid $^{4}$He
from a normal phase to a superfluid one at a finite rate. 
Suppose the transition point is crossed at time $t=0$,
while  time evolution starts at $t\ll0$.
As long as the liquid is far away from  the transition point 
its time evolution is adiabatic. 
In other words, the relaxation time scale $\tau$,
which tells how much time the system needs to 
adjust to new thermodynamic conditions, is small enough. 
As the transition is approached
the critical slowing down occurs, i. e. $\tau\to\infty$,
so that at the instant $-\hat{t}$ 
the system leaves adiabatic regime and enters an impulse one where
its state is effectively  frozen-- see Fig. \ref{fig1}a for an illustration 
of these concepts. 
The time $\hat{t}$ is called the freeze-out time and was introduced
by Zurek \cite{zurek}.  
As the quench proceeds after crossing the transition point, 
the relaxation time scale decreases. At the instant $\hat{t}$,
the system goes back into an adiabatic regime.
The freeze-out time is determined by the Zurek's 
equation: $\tau(\hat{t})=\hat{t}$ \cite{zurek}. For the case of liquid 
$^4$He it was found experimentally 
that $\tau=\tau_0/|\varepsilon|$, where $\tau_0$ is a constant,
while $\varepsilon$ is called the relative temperature. The latter 
measures the distance of the liquid from a  transition point being 
at $\varepsilon=0$, i. e. $\varepsilon(t=0)=0$. 
Physically changes of pressure translate into
changes of  $\varepsilon$. It is further assumed that pressure changes are such that 
$\varepsilon=t/\tau_Q$, where $\tau_Q$ is a quench timescale.
Now the Zurek's equation reads: $\tau_0\tau_Q/\hat{t}=\hat{t}$, which 
results in $\hat{t}=\sqrt{\tau_Q\tau_0}$.
As shown in \cite{zurek}, knowledge of $\hat{t}$ allows for making a 
prediction 
of density of topological defects, resulting from a nonequilibrium 
phase transition, without 
solving dynamical equations describing the system!

We consider time dependent  Hamiltonian 
\begin{equation}
\label{eq1}
\frac{1}{2}
\left(
\begin{array}{cc}
\Delta\cdot t & \omega_0 \\
\omega_0 & -\Delta\cdot t
\end{array}
\right)
\end{equation}
written in the  basis of time independent 
states $|1\rangle$ and $|2\rangle$. Eigenstates of (\ref{eq1}) have the form 
$$
\left[
\begin{array}{c}
|\uparrow(t)\rangle \\ |\downarrow(t)\rangle 
\end{array}
\right] = 
\left(
\begin{array}{rr}
\cos(\theta(t)/2)  & \sin(\theta(t)/2) \\
-\sin(\theta(t)/2) & \cos(\theta(t)/2)
\end{array}
\right)
\left[
\begin{array}{c}
|1\rangle \\ |2\rangle 
\end{array}
\right],
$$
where $\cos(\theta)=\varepsilon/\sqrt{1+\varepsilon^2}$, 
$\sin(\theta)=1/\sqrt{1+\varepsilon^2}$, 
$\theta\in[0,\pi]$,  
$\varepsilon= \Delta\cdot t/\omega_0$.
As in   LZ theory $\Delta, \omega_0>0$ 
are constant parameters.
The level structure of (\ref{eq1}) is depicted in  lower part of 
Fig. \ref{fig1},
while the gap equals $\sqrt{\omega_0^2+(\Delta\cdot t)^2}$. 

Topological defects can be introduced into the 
LZ model in the following way.
Suppose  the state $|1\rangle$ corresponds to a vortex state being 
an eigenstate of angular momentum operator:  $\hat{L}_z|1\rangle=n|1\rangle$ 
($n=\pm1,\pm2,\dots$), while the state $|2\rangle$ satisfies 
$\hat{L}_z|2\rangle=0$.
System's wave function can be written as 
$|\Psi\rangle= a |1\rangle+ b|2\rangle$ ($\protect{|a|^2+|b|^2=1}$, 
$\langle i|j \rangle=\delta_{ij}$). 
We propose to identify the density of topological defects with 
the average value of angular momentum: 
$\protect{\langle\Psi|\hat{L}_z|\Psi\rangle= n |a|^2}$ \cite{vortex}. 
For the rest of discussion we  define normalized to unity 
density of defects as \cite{definicja}
\begin{equation}
\label{D}
{\cal D}_n:= \langle\Psi|\hat{L}_z|\Psi\rangle/n
= |\langle\Psi|1\rangle|^2.
\end{equation}
Suppose now that the system undergoes adiabatic time evolution from the
ground state of (\ref{eq1}) at $t\to-\infty$ to 
the ground state of (\ref{eq1}) at $t\to\infty$. 
Therefore, the state of the system undergoes the ''phase transition''
from $|1\rangle$ to $|2\rangle$, i. e.
from a vortex-defected ''phase'' to a vortex-free one. 
If  time evolution fails to be adiabatic, which is usually the case,
the final state of the system is a superposition of states $|1\rangle$ and
$|2\rangle$ so that the final
density of topological defects becomes non-zero.
We will show that the KZ-like theory predicts surprisingly  
correctly vortex density (\ref{D}) as a function of a
transition rate only.

Analogs of relaxation timescale, relative temperature and quench timescale
are identified as follows. First of all, let us recall 
ingenious  
simplification of system's dynamics in the KZ theory.
The simplification relies on the assumption that the system either
evolves adiabatically, i. e. adjusts perfectly to changes of parameters,
or becomes immobilized, i. e. undergoes the so called impulse evolution
\cite{zurek}. As proposed by Zurek, the switch between 
adiabatic/impulse regimes is determined by the relaxation time scale,
which is small during an adiabatic evolution and large in the
impulse part.
We would like to employ similar strategy  below. 
From the adiabatic theorem one knows that 
as long as the inverse of the gap is small enough 
the system starting evolution from a ground state 
remains in  the ground state. It naturally suggests that 
inverse of the gap, being necessarily small in the adiabatic part of evolution,
can be considered as  a
quantum mechanical equivalent of the relaxation timescale
introduced above: we set $\tau=1/\sqrt{\omega_0^2+(\Delta\cdot t)^2}$ 
\cite{relaxation}. 
The equivalent
of the relative temperature $\varepsilon$, i. e. a dimensionless
distance of the system from anticrossing,
is  $\Delta\cdot t/\omega_0$. As a 
quench timescale $\tau_Q$ we take  $\omega_0/\Delta$, while 
$\omega_0$ we identify with $1/\tau_0$.
Finally, we arrive at 
\begin{equation}
\label{tau}
\tau=\frac{\tau_0}{\sqrt{1+\varepsilon^2}} \ \ \ , \ \ \
\varepsilon=\frac{t}{\tau_Q}.
\end{equation}
For $|\varepsilon|\gg1$ expressions (\ref{tau}) 
are identical as those introduced above
in the context of topological defect production in liquid $^4$He,
which will be commented  below.

In the following we 
consider dynamics of the LZ model described by the Schr\"odinger 
equation: 
$i \frac{d}{dt}|\Psi\rangle=\hat{H}|\Psi\rangle$, 
with $\hat{H}$ given by (\ref{eq1}). 
We assume that 
time evolution starts from a ground state of (\ref{eq1})
at some $t=t_i$ and 
lasts till $t_f\to+\infty$. The quantity of interest will be 
density of defects (\ref{D}) at the end of time evolution, which is in fact 
the probability of finding the system in the excited eigenstate 
at $t_f$.
Adopting the KZ simplification of the system's dynamics,
we assume that the evolution of the system is either adiabatic or diabatic.
The adiabatic part takes place when the system is away from the
anti-crossing, while the diabatic part takes place in the neighborhood of an anticrossing,
where the inverse of the gap is so large that the system no longer adjusts to the changes 
of the
Hamiltonian (Fig. \ref{fig1}b)- compare to the pressure induced quench in
$^4$He described above. 
Therefore, the two non-trivial schemes can be considered:
\begin{itemize}
\item[\bf A:] $t_i<-\hat{t}$: the evolution starts in the adiabatic regime, so it 
is adiabatic from $t_i$ till $-\hat{t}$, 
then impulse from $-\hat{t}$ to $\hat{t}$, 
and finally adiabatic from $\hat{t}$ to $t_f$-- see Fig. \ref{fig1}b, 
\item[\bf B:] $t_i\in[-\hat{t},\hat{t}\,]$: the evolution starts
in the impulse regime, therefore it is 
impulse from  $t_i$ to $\hat{t}$ and then adiabatic from $\hat{t}$ to $t_f$
-- compare to Fig. \ref{fig1}b.
\end{itemize}
The statement that the evolution is impulse
means that the system's wave function
changes by the overall phase factor only. Additionally, it is assumed that
we do not consider slow time evolutions for which the system stays
whole time in the adiabatic regime due to the finite gap of the LZ model.
Notice that such  evolutions would be incompatible with
KZ considerations where divergence of relaxation time scale at the transition
point, Fig. \ref{fig1}a,  prohibits adiabatic evolutions close to the transition point.
The assumptions standing behind A (B) scheme classification are
approximate and heuristic as the whole KZ theory is,
and our aim is to find how good they work in the LZ system.

The only quantity that is still unknown is the instant $\hat{t}$.
It is found from the equation originally proposed
by Zurek in the context of classical phase transitions \cite{zurek}

\begin{equation}
\label{freeze}
\tau( \hat{t}\, )=\alpha\hat{t},
\end{equation}
and modified by us by a factor $\alpha={\cal O}(1)$, i. e. 
the only free parameter of our theory {\it independent}
of $\tau_Q$ and    $\tau_0$. 
The solution of (\ref{freeze}) reads
\begin{equation}
\hat{\varepsilon}=\varepsilon(\hat{t}\,)=
\frac{1}{\sqrt{2}}\sqrt{\sqrt{1+\frac{4}{x_\alpha^2}}-1}, 
\ \ \ x_\alpha=\alpha
\frac{\tau_Q}{\tau_0}.
\label{hatt}
\end{equation}
The first observation shows that for fast transitions, i. e. $x_\alpha\to0$ 
at $\tau_0$ being fixed, one gets  
$\hat{t}=\sqrt{\tau_0\tau_Q/\alpha}$. 
Therefore we recover, up to ${\cal O}(1)$ 
factor, the well known result \cite{zurek}. It happens because 
in the fast transition limit $\hat{\varepsilon}\gg1$ and then 
$\tau(\hat{\varepsilon})\approx \tau_0/\hat{\varepsilon}$, which is
the same as in the theory of dynamics of quantum phase transitions
in liquid $^4$He \cite{zurek}. 
This observation further supports similarities of our model to 
KZ systems.

For the first application of our theory we consider the situation when time
evolution starts far away from the anticrossing 
- a generic {\bf A} scheme case. 
Taking $|\Psi(t_i)\rangle=|\downarrow(t_i)\rangle$ as an initial system's wave function,
and assuming limits $t_i\to-\infty$ and  $t_f\to\infty$,
one easily gets the following final density of topological 
defects 
\begin{equation}
\label{standardlz}
{\cal D}_n= |\langle \Psi(t_f)|1\rangle|^2\approx 
|\langle \uparrow(\hat{t}\,)|\downarrow(-\hat{t}\,)\rangle|^2= 
\frac{\hat{\varepsilon}^2}{1+\hat{\varepsilon}^2}.
\end{equation}
Derivation of (\ref{standardlz}) uses  the following relations:   
$|\protect{\langle \uparrow(t_f)|\Psi(t_f)\rangle}|\approx
|\protect{\langle \uparrow(\hat{t}\,)|\Psi(\hat{t}\,)\rangle}|
\approx |\protect{\langle\uparrow(\hat{t}\,)|\Psi(-\hat{t}\,)\rangle}|
\approx |\protect{\langle\uparrow(\hat{t}\,)|\downarrow(-\hat{t}\,)\rangle}|
$.
Substitution of (\ref{hatt}) into (\ref{standardlz}) gives  
\begin{equation}
\label{P}
{\cal D}_n= \frac{2}{{\cal P}(x_\alpha)} \ \ \ , \ \ \
{\cal P}(x_\alpha)= x_\alpha^2+ x_\alpha\sqrt{x_\alpha^2+4}+2.
\end{equation}
Expanding ${\cal D}_n$ into a series one gets for fast transitions 
\begin{equation}
\label{1}
{\cal D}_n= \exp(-x_\alpha)+ {\cal O}\left(x_\alpha^3\right),
\end{equation}
which is an exact result up to 
${\cal O}\left(x_\alpha^3\right)$ terms if the constant $\alpha$ 
is chosen as
$\pi/2$ \cite{LZ}. Notice that $\alpha={\cal O}(1)$ as assumed in 
(\ref{freeze}). In the adiabatic limit ($x_\alpha\to\infty$),
Eq. (\ref{P}) predicts 
${\cal D}_n={\cal O}(1/x_\alpha^2)$ instead of exponential decay, 
which does not affect  results much due to 
 very small value of ${\cal D}_n$ in that
regime. 

The best performance for fast transitions can be understood as
follows. 
The derivation of (\ref{standardlz}) requires assumption that in the time
interval $[-\hat{t},\hat{t}\,]$ the state of the system does not change essentially. 
The smaller is this 
time interval the better is this assumption. 
From (\ref{tau}) and (\ref{hatt})  one easily 
finds that $\hat{t}/\tau_0$ 
grows monotonically with $x_\alpha$. Indeed,  
$\hat{t}/\tau_0$ equals $\sqrt{x_\alpha}/\alpha$ for $x_\alpha\to0$ 
and increases
to $1/\alpha$ for $x_\alpha\to\infty$. Therefore it is not surprising that 
our predictions work better for fast transitions.

\begin{figure}
\includegraphics[angle=-0,scale=0.25, clip=true]{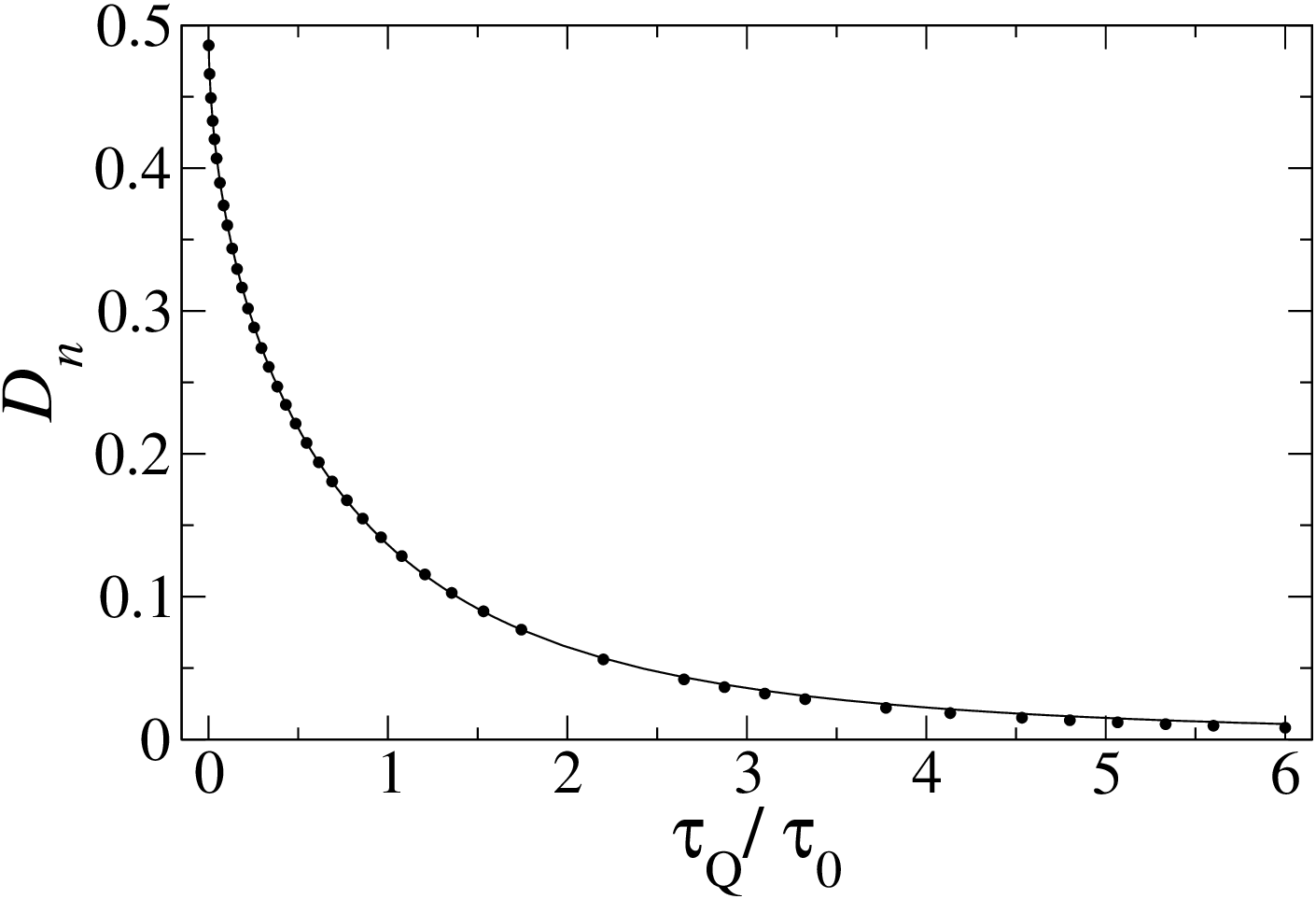}
\caption{Density of defects for the system whose evolution 
starts at the anticrossing center.
Solid line-  prediction
(\ref{closed}), dots- numerical data. The parameter 
$\alpha=0.77$ was found from  fit of (\ref{closed}) to numerics.} 
\label{fig2}
\end{figure}

Now we would like to discuss the situation when
time evolution starts from a ground state at the anticrossing center, $t_i= 0$, which 
is a generic {\bf B} scheme situation. As $t_f\to\infty$ one gets  
\begin{equation}
\label{second1}
{\cal D}_n = |\langle \uparrow(\hat{t}\,)|\downarrow(0)\rangle|^2=
 \frac{1}{2}\left(1-\frac{1}{\sqrt{1+\hat{\varepsilon}^2}}\right),
\end{equation}
where we put $|\Psi(0)\rangle=|\downarrow(0)\rangle=-\frac{\sqrt{2}}{2}
|1\rangle + \frac{\sqrt{2}}{2}|2\rangle$,  and 
assumed that  
$|\protect{\langle \uparrow(t_f)|\Psi(t_f)\rangle}|\approx
\protect{|\langle\uparrow(\hat{t})|\Psi(\hat{t}\,)\rangle}|\approx
|\protect{\langle \uparrow(\hat{t}\,)|\Psi(0)\rangle}|$.
Combining (\ref{hatt}) and (\ref{second1}) one gets 
\begin{eqnarray}
\label{closed}
{\cal D}_n= \frac{1}{2}\left(1-\sqrt{1-2/{\cal P}(x_\alpha)} \right),
\end{eqnarray}
with $x_\alpha$ and ${\cal P}(x_\alpha)$ defined in (\ref{hatt}) and (\ref{P}). 
The agreement of this expression with results of numerical calculations
is remarkable as depicted in Fig. \ref{fig2}. It is even 
better than in the previous case when we considered 
the evolution starting far away from the avoided crossing.
We attribute it to the fact that now the frozen part  
takes less time, i. e. $\hat{t}$ instead of $2\hat{t}$, 
and to the absence of the approximation that 
the initial stage of evolution is adiabatic.

\begin{figure}
\includegraphics[angle=-0,scale=0.25, clip=true]{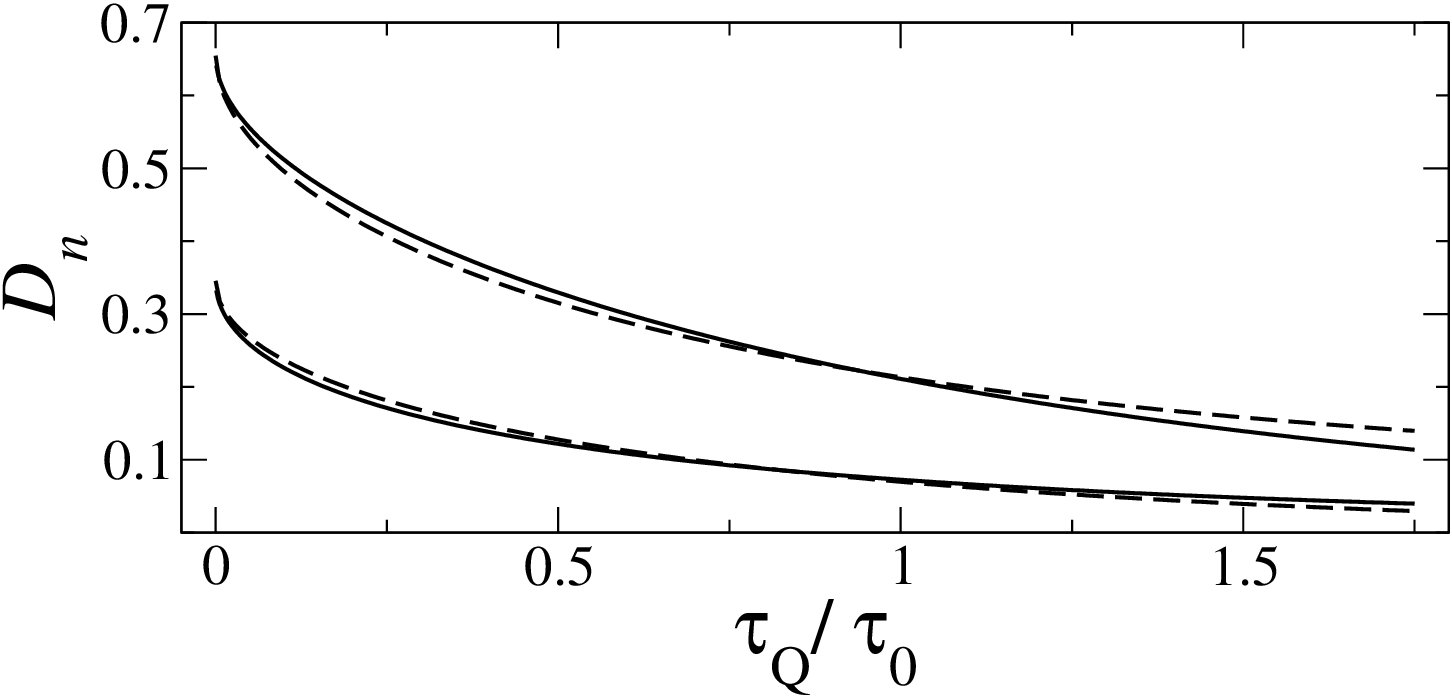}
\caption{Density of defects. Solid lines- numerics, dashed lines- 
Eq. (\ref{theta}) with $\alpha$ determined from fit. 
Upper (lower) curves correspond 
to $\theta_0= 0.6\pi$, $\alpha=1.06$ ($\theta_0= 0.4\pi$,
$\alpha=0.58$).} 
\label{fig3}
\end{figure}

We can also easily calculate 
density of defects when time evolution starts 
in the impulse regime ({\bf B} scheme),  but
outside the avoided crossing center.
Taking $|\Psi(t_i)\rangle=|\downarrow(t_i)\rangle:=-\sin(\theta_0/2)|1\rangle
+\cos(\theta_0/2)|2\rangle$ we obtained 
\begin{eqnarray}
\label{theta}
{\cal D}_n= -\frac{\cos(\theta_0)}{\sqrt{2{\cal P}(x_\alpha)}}
+\frac{1-\sqrt{1-2/{\cal P}(x_\alpha)}\sin(\theta_0)}{2},
\end{eqnarray}
where $\theta_0=\arctan(\omega_0/(\Delta\cdot t_i))\in[0,\pi]$ 
measures  distance of the starting point of time evolution 
from an avoided crossing,
e. g. $\theta_0=\pi/2$
when evolution starts from an anticrossing center and then
Eq. (\ref{theta}) is the same as Eq. (\ref{closed}).

Comparison of (\ref{theta}) to numerics  
for $\tau_Q/\tau_0\le1.75$ and 
$|\theta_0-\pi/2|\le \pi/10$ 
reveals  satisfactory agreement- see Fig. \ref{fig3} for 
a typical situation.  
For larger $\tau_Q/\tau_0$ and/or $|\theta_0-\pi/2|$ the 
agreement gradually decreases, which 
we attribute to the fact that for these parameters 
the starting time moment, $t_i=\tau_Q/\tan\theta_0$,
might be outside $[-\hat{t},\hat{t}\,]$,
so that the assumption that the initial stage of time evolution is 
impulse can be wrong. One avoids these problems  
when either $t_i\ll-\hat{t}$ or $|t_i|\ll\hat{t}$, i. e. 
when the system evolves clearly within the {\bf A} or {\bf B} 
scheme, respectively. 

Having at hand above obtained results, let us comment on the
the Zurek-like equality (\ref{freeze}) extensively used in this paper.
This equality gives the time moments $\pm\hat{t}(\tau_Q)$, which 
separate the adiabatic and impulse regimes \cite{fixed}. 
Do we need to rely on this equation?
To answer this question we notice that 
we aim at getting the best description of LZ model 
dynamics by using the simplification that the evolution 
is either adiabatic or impulse in the sense specified 
below the A (B) scheme description. It means that the whole
problem can be reduced to getting the time moments $\pm\hat{t}$
that lead to the best comparison of defect density to 
exact results. This can be done without 
Eq. (\ref{freeze}) by fitting the 
$\hat{t}(\tau_Q)$ directly to numerical data.
Eq. (\ref{freeze}) reduces  the problem of getting
optimal $\hat{t}(\tau_Q)$ to fitting just one parameter, $\alpha$, 
to the numerical  results instead of fitting a whole function 
$\hat{t}(\tau_Q)$. Naturally, it is of fundamental interest to 
try to find analytically the best  $\hat{t}(\tau_Q)$ function
and compare it to those taken directly from the KZ theory.

Let us point out possible usefulness of these considerations
in understanding of  non-equilibrium
quantum phase transitions (QPTs).
There is no doubt that  quantum many-body systems of interest
(spin systems, cold atoms in optical lattices) are more
complicated than the LZ model. It is, however,
generally accepted that close to the generic second order QPT point
there exists an
anti-crossing between a ground state and a first excited state-
see Ch. 1.1 of \cite{subir} and Sec. 2.4 of \cite{vojta}.
Therefore, one can  expect that
at least a qualitative picture of the change of system's
properties during  a  QPT can be captured by a simple two level model,
e. g. the LZ model. Notice that if it would happen that another sort of
a two level approximation would work better, still it is quite likely that
the same analysis as the one presented here would work.
Once the many-body model of interest is
specified and its low energy static properties are determined, one can
define an equivalent of the  "density of defects" and study system's
dynamics using the tools presented in this Letter.
In fact, the work, along the lines of this paper,
on a dynamics of the  superfluid-Mott insulator
QPT of cold atoms in an optical lattice
is in progress now.

Let's look at other possible extensions of this work.
First of all, it is desired to re-analyze more strictly the 
LZ dynamics, to get a 
systematic control of the intuitive results obtained above.
This work can be done utilizing the results from \cite{vitanow}.
Second, it seems to be very interesting to 
investigate how the adiabatic/impulse simplification of 
system dynamics works in other quantum mechanical systems, for instance
those which exhibit faster increase of the gap with the distance 
from anti-crossing. We expect to get better qualitative agreement in these
cases.

Several other remarks are in order. First of all, we have 
shown that the very simple LZ model successfully reproduces
KZ-like dependence of topological defect density on the quench rate.
Second, our results provide intuitive description of LZ model dynamics
unexplored up to now  in numerous papers devoted to
the LZ model. Notice that with our intuitive approach we have been able 
to obtain qualitatively correct predictions concerning LZ model dynamics {\it without}
solving the time-dependent Schr\"odinger equation!
Third, we show that the KZ theory can provide
predictions beyond the lowest non-trivial 
$\tau_Q/\tau_0$ order usually considered \cite{zurek,siec,antunes}.
Fourth, our results are directly applicable 
to different
quantum two level systems, e. g. to the molecular magnets Fe$_8$ 
\cite{magnets}. In particular, we expect that 
our prediction that  the transition probability
for fast quenches (small $\tau_Q/\tau_0$) decays in a power-like manner
instead of an exponential one, Eq.  (\ref{closed}),  
can be experimentally verified and might be helpful in
interpretation of experimental data. 

I would like to thank Jacek Dziarmaga
and Wojciech Zurek for discussions. 
Financial support from the Alexander von Humboldt Foundation and 
DFG (SFB 407) is also gratefully acknowledged.


\begin{thebibliography}{99}
\bibitem{definicja} 
Considering the freedom in choosing
the states $|1\rangle, |2\rangle$, it is clear that
other, non-vortex based  definitions of defect densities,
can be worked out
when a particular form of wave functions is assumed.
\bibitem{relaxation} Other definitions of $\tau$ 
seem to be equally plausible on the heuristic 
level considered here,
e. g. 
$\tau=\omega_0/[\omega_0^2+(\Delta\cdot t)^2]\sim 1/{\rm gap}^2$. 
We  studied a few such choices of $\tau$  and found that
the simplest one, $\tau\sim1/{\rm gap}$ discussed in the Letter, 
compares the best to exact results.
\bibitem{fixed}
Notice that $\hat{t}(\tau_Q)$ provides Zurek's freeze-out time 
for different transitions labeled by $\tau_Q$,
but all of them  starting from the same initial (ground) state, 
e. g. from the ground state at the center of the anti-crossing.
\end{thebibliography}
\end{document}